\documentclass[aps,pra,reprint,hidelinks,nobalancelastpage,longbibliography]{revtex4-1}
\usepackage{amsmath,braket}
\usepackage{graphicx}
\usepackage{color}
\usepackage{hyperref}
\usepackage[separate-uncertainty=true,multi-part-units=single]{siunitx}

\begin{document}
\title{Phonon impact on optical control schemes of quantum dots: The role of quantum dot geometry and symmetry}

\author{S.~L\"uker}
\affiliation{Institut f\"ur Festk\"orpertheorie, Universit\"at M\"unster,
Wilhelm-Klemm-Str.~10, 48149 M\"unster, Germany}

\author{T.~Kuhn}
\affiliation{Institut f\"ur Festk\"orpertheorie, Universit\"at M\"unster,
Wilhelm-Klemm-Str.~10, 48149 M\"unster, Germany}

\author{D.~E.~Reiter}
\affiliation{Institut f\"ur Festk\"orpertheorie, Universit\"at M\"unster,
Wilhelm-Klemm-Str.~10, 48149 M\"unster, Germany}

\date{\today}

\begin{abstract}
Phonons strongly influence the optical control of semiconductor quantum dots.
When modeling the electron-phonon interaction in several theoretical
approaches the quantum dot geometry is approximated by a spherical structure, though typical self-assembled quantum dots are strongly lens-shaped. By
explicitly comparing simulations of a spherical and a lens-shaped dot using a
well-established correlation expansion approach we show that indeed
lens-shaped dots can be exactly mapped to a spherical geometry when studying
the phonon influence on the electronic system. We also give a recipe to
reproduce spectral densities from more involved dots by rather simple
spherical models. On the other hand, breaking the spherical symmetry has a
pronounced impact on the spatio-temporal properties of the phonon dynamics.
As an example we show that for a lens-shaped quantum dot the phonon emission
is strongly concentrated along the direction of the smallest axis of the dot
which is important for the use of phonons for the communication between
different dots.
\end{abstract}



\maketitle

\section{Introduction}
In the optical control of semiconductor quantum dots (QDs) the coupling to
phonons plays a vital role \cite{reiter2014the,ramsay2010are}. Nowadays the
optical control of excitons and biexcitons in single QDs can be performed
experimentally by various techniques and the results clearly demonstrate the
influence of phonons. Examples are the phonon-induced damping of Rabi
rotations in the case of resonant driving of the QD \cite{ramsay2010dam},
chirp-dependent phonon damping of the adiabatic rapid passage
induced by chirped laser pulses
\cite{mathew2014sub,kaldewey2017coh,kaldewey2017dem}, phonon-assisted state
preparation by detuned excitation
\cite{quilter2015pho,liu2016ult,bounouar2015pho,ardelt2016opt,reindl2017pho}
and the polarization decay in four-wave mixing signals
\cite{jakubczyk2016imp,wigger2017exp}. Also for QDs coupled to a quantized
cavity light field phonons play an important role
\cite{hughes2011inf,ulhaq2013det,kaer2013mic,gustin2017inf,grange2017red}. To theoretically
simulate the phonon influence different theoretical approaches have been
developed, identifying the coupling to longitudinal acoustic (LA) phonons as
typically being the dominant mechanism. Without optical excitation, the
system can be described by the independent Boson model, which is exactly
solvable. In the phonon-free case the optically driven QD can also be exactly
solved using the semiconductor Bloch equations. When both light field and
phonons are active, an exact solution cannot be found anymore and
approximations or numerical methods have been developed. Examples are
different types of master-equation approaches
\cite{mccutcheon2011gen,manson2016pol,nazir2016mod,reiter2017tim},
time-convolutionless methods \cite{machnikowski2008the,gawarecki2012dep},
correlation expansion techniques
\cite{forstner2005pho,krugel2005the,luker2012inf,reiter2012pho} and path integral
formulations \cite{vagov2011rea,glassl2011lon,glassl2011inf}. The methods
rely on different levels of approximations coming with different advantages
and disadvantages. In particular, the computational costs of the different
methods can greatly vary. In order to reduce the computational costs, in many
cases the approximation of a spherical QD has been made. However, fabrication
techniques like, e.g., the Stranski-Krastanov growth mode for self-assembled
QDs typically lead to rather lens-shaped geometries, as has also been
confirmed by many experimental measurements \cite{woggon1997opt,fry2000pho}.
Nevertheless, various recent combined experimental-theoretical studies on
single QDs, in which the theoretical calculations have been based on a
spherical dot approximation, have shown an excellent agreement between theory
and experiment
\cite{ramsay2010dam,quilter2015pho,kaldewey2017coh,kaldewey2017dem}. In this
paper, we analyze the impact of the approximation of spherical symmetry for
simulations of the phonon influence on the dynamics of optically excited QDs.

For a spherical QD, the electron-phonon coupling matrix element only depends
on the modulus of the phonon wave vector, which allows one to exploit
symmetry arguments to reduce the number of independent variables and thereby
the computational cost. In contrast, for a lens-shaped QD, at least two
directions have to be taken into account, which often increases the number of
variables and the computational cost considerably. In this paper we provide a
direct comparison between a lens-shaped and a spherical QD using the same
approximations in the theoretical model. To be specific, we here explicitly
compare calculations performed for a spherical and a lens-shaped QD within a
correlation expansion \cite{krugel2005the,luker2012inf,glassl2011inf} and
show that the optically induced electronic properties of the lens-shaped
system are perfectly reproduced by a model involving spherical symmetry. As
an example we compare the occupation under resonant and chirped excitation.
We furthermore show that for each spectral density we can find a spherically
symmetric effective confinement potential, which reproduces the spectral
density resulting in the same effects on the electronic properties. In
addition we show that the same spectral density can be well approximated by
rather simple spherical models with only very few parameters.

On the other hand, when considering the phonon system, the symmetry
properties of the QD have a great impact. Using the correlation expansion we
calculate the dynamics of the phonons created during and immediately after
the optical excitation \cite{wigger2013flu,wigger2014ene}. The rapid
generation of the exciton leads to the formation of a wave packet leaving the
QD region \cite{vanacore2017ul,jakubczyk2016imp}. For a spherical QD, the
generated wave packet retains the spherical symmetry of the QD. For a
lens-shaped QD the wave packet generation reflects the QD symmetry
\cite{krummheuer2005pur}, the strongest wave packet emission being in the
direction of the smallest size.

\section{Theoretical background}
To study the influence of phonons on the QD exciton system, we make use of
the standard two-level model of a strongly confined QD consisting of the
ground state $|g\rangle$ at energy $0$ and the exciton state $|x\rangle$ at
energy $\hbar\omega$ \cite{luker2012inf,reiter2014the}. The Hamiltonian for
the electronic part of the system can be written as
\begin{equation}
	H_{el}= \hbar\omega |x\rangle \langle x| - \mathbf{E}\cdot\mathbf{P}.
\end{equation}
The second term describes the coupling of the polarization
\begin{equation}
	\mathbf{P}= \mathbf{M}|g\rangle \langle x|+\mathbf{M}^*|x\rangle \langle g|,
\end{equation}
$\mathbf{M}$ being the dipole matrix element, to the light field
$\mathbf{E}$, which is treated in dipole and rotating wave approximation.

The phonon part of the system consists of the free phonon part and the
electron-phonon interaction described by
\begin{equation}
	H_{ph} = \hbar\sum_{\mathbf{q}}\omega_{\mathbf{q}}b^\dag_{\mathbf{q}}b_{\mathbf{q}}+
\hbar |x\rangle\langle x| \sum_{\mathbf{q}}\left(g_{\mathbf{q}}b^{}_{\mathbf{q}} +
g_{\mathbf{q}}^*b_{\mathbf{q}}^\dag\right),
\end{equation}
with $b_{\mathbf{q}}$ ($b^\dag_{\mathbf{q}}$) being the annihilation
(creation) operator for a phonon with wave vector $\mathbf{q}$. We consider
LA phonons with linear dispersion $\omega_{\mathbf{q}} = c_{LA}q$ ($c_{LA}$
being the sound velocity).
$g_{\mathbf{q}}=g_{\mathbf{q}}^{e}-g_{\mathbf{q}}^{h}$ is the coupling matrix
element for deformation potential coupling given by
\begin{equation}\label{eq:ep-coupling}
	g_{\mathbf{q}}^{e/h}=\sqrt{\frac{q}{2V\rho\hbar c_{LA}}}D_{e/h}F_{\mathbf{q}}^{e/h}
\end{equation}
for electrons (e) and holes (h). The constants are the normalization volume
$V$, the mass  density $\rho$, and the deformation potential coupling
constant $D_{e/h}$ for electrons/holes. The spatial confinement of the
carriers enters via the form factors
\begin{equation} \label{eq:ff_wavefunction}
	F_{\mathbf{q}}^{e/h}=\int d^3r \,|\psi^{e/h}(\mathbf{r})|^2 \, e^{i\mathbf{q}\cdot\mathbf{r}}
\end{equation}
depending on the envelope function $\psi^{e/h}(\mathbf{r})$ of
electrons/holes. For the QD, we assume a harmonic confinement potential.
Considering a lens-shaped QD with the localizations length $a^{e/h}_z$ in the
$z$-direction and $a^{e/h}_r$ in the $(x,y)$-plane (with
$a^{e/h}_z<a^{e/h}_r$), the form factor reads
\begin{equation} \label{eq:ff_lens}
	F_{\mathbf{q}}^{e/h}=\exp\left[-\frac{1}{4}\left(q_z^2
(a^{e/h}_z)^2+q_{r}^2(a^{e/h}_r)^2\right)\right],
\end{equation}
where $q_z$ is the wave vector component in $z$-direction, $q_{r}$ is the
modulus of the in-plane wave vector and $q=\sqrt{q_r^2+q_z^2}$. In the case
of a spherical QD we have $a^{e/h}_{z}=a^{e/h}_{r}=a^{e/h}$ simplifying the
form factor to
\begin{equation} \label{eq:ff_sphere}
	F_{\mathbf{q}}^{e/h}=\exp\left[-\frac{1}{4}q^2 (a^{e/h})^2\right].
\end{equation}
Here we already see that the numerical implementation of a spherical QD
relies  only on the modulus of the wave vector $q$, while for a lens-shaped
QD a discretization of both $q_z$ and $q_r$ is required. If not denoted
otherwise, we use standard GaAs parameters (see table
\ref{material-parameters}) which, assuming the same confinement potential for
electrons and holes, give the ratio of the localization length between
electron and hole to $a^h/a^e=(m_e/m_h)^{1/4}=0.87$ \cite{krugel2006bac}.

Using the Hamiltonian given above the equations of motion are set up. In the
density matrix formalism, starting from the basic electronic and phononic
variables exciton occupation $f=\langle |x\rangle \langle x| \rangle$,
excitonic polarization $p=\langle |g\rangle \langle x| \rangle$, coherent
phonon amplitudes $B_{\mathbf{q}}=\langle b_{\mathbf{q}}^{} \rangle$ and
phonon occupations $n_{\mathbf{q},\mathbf{q'}} = \langle b_{\mathbf{q}}^\dag
b_{\mathbf{q'}}^{} \rangle$, the many body nature of the carrier-phonon
coupling leads to an infinite hierarchy of equations of motion. In the higher
orders, phonon-assisted quantities appear like $\langle b_{\mathbf{q}}^\dag
|x\rangle \langle x|  \rangle$ or $\langle b_{\mathbf{q}}^\dag
b_{\mathbf{q}}^\dag |x\rangle \langle x| \rangle$. In the correlation
expansion the correlations of the phonon-assisted quantities are neglected at
a certain order. Obviously, the dimension of $\mathbf{q}$ is directly related
to the dimension of the phonon-assisted quantities. Hence, the restriction to
a one-dimensional $\mathbf{q}$ goes along with an immense saving of
computational cost. In the following we will employ a fourth-order
correlation expansion, which has proven to simulate experimental results very
well \cite{kaldewey2017coh,kaldewey2017dem,reiter2014the}.

\begin{table}
\caption{Material parameters}\label{material-parameters}
\begin{tabular}{l l l}
\hline
Longitudinal sound velocity & $c_{LA}$ & $5.1\,$nm ps$^{-1}$\\
Mass density & $\rho$ & $5.31\,$g cm$^{-3}$\\
Electron deformation potential & $D_e$ & $7\,$eV\\
Hole deformation potential & $D_h$ & $-3.5\,$eV\\
Effective electron mass & $m_e$ & $0.067 m_0$\\
Effective hole mass & $m_h$ & $0.110 m_0$\\
Free electron mass & $m_0$ & $9.1 \cdot 10^{-31}\,$kg\\
\hline
\end{tabular}
\end{table}

\section{Spectral density}
The phonon spectral density provides a measure for the coupling efficiency
between the phonons and the carriers. It is defined as
\begin{equation} \label{eq:phonon-spect}
 J(\omega) = \sum_{\bf q} |g_{\bf q}|^2 \delta(\omega-\omega_{\bf q}) \, .
\end{equation}
For a spherical QD with $a^e = a^h = a$ an analytical expression for the
spectral density is available \cite{calarco2003spi,mccutcheon2010qua} which
has the form
\begin{equation}
 J(\omega) = A \omega^3 \exp(-\omega^2/\omega_c^2),
\label{eq:phonon-spect-anal}
\end{equation}
where $\omega_c$ is the cut-off frequency determined by the size of the QD
and $A$ is a measure for the coupling strength. For the coupling given in
Eq.~(\ref{eq:ep-coupling}) these parameters are given by
\begin{equation}
 A = \frac{(D_e-D_h)^2}{4\pi^2\rho\hbar c_{LA}^{5}} , \quad \omega_c= \sqrt{2}\frac{c_{LA}}{a}.\notag
\end{equation}
\begin{figure}[t]
	\includegraphics[width=\columnwidth]{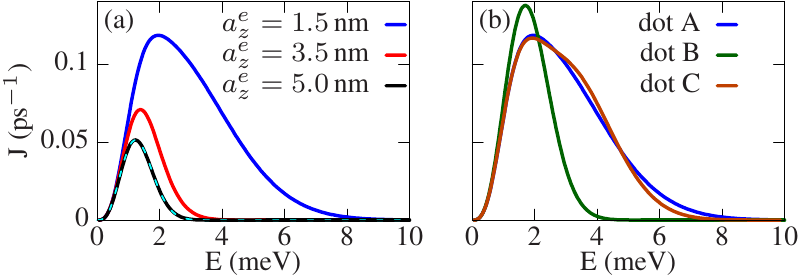}
	\caption{\label{fig:spectraldensity} Phonon spectral density for (a) a QD
with a in-plane size $a^e_r=5.0$~nm and different heights $a^e_z=1.5, 3.5$ and
$5.0$~nm. The dashed line is the spectral density according to Eq.~\eqref{eq:phonon-spect-anal}.
(b) Phonon spectral density for QDs A, B and C as explained in the text.}
\end{figure}
The spectral densities of differently shaped QDs are compared in
Fig.~\ref{fig:spectraldensity}. In Fig.~\ref{fig:spectraldensity}(a) we keep
the in-plane localization length of the electrons fixed to $a^e_r=5.0$~nm and
vary the size in $z$-direction $a^e_z$. The corresponding localization
lengths of the holes are taken to be $a^h=0.87 a^e$. All curves exhibit a
non-monotonic behavior with a pronounced maximum. Phonons with energies $E =
\hbar\omega_{\mathbf{q}}$ in the region around the maximum bear the major
impact on the carrier dynamics. For $a^e_z=5$~nm, the QD is spherical and the
maximum is at about $1.2$~meV. The dashed curve shows the analytical equation
\eqref{eq:phonon-spect-anal}, the fit parameters are $A = 0.036\,$ps$^2$ and
$\omega_c = 1.52\,$ps$^{-1}$. Though the considered dot does not
fulfill $a^e = a^h$, the analytical formula leads to an excellent agreement
with the phonon spectral density of the spherical QD. The cut-off frequency
of the fit corresponds to a localization length of $4.75$~nm, which is
between the values for electrons ($5$~nm) and holes ($4.35$~nm). With
decreasing height the QD becomes more lens-shaped. In the spectral density
the maximum shifts to higher energies with $1.4$~meV for $a^e_z=3.5$~nm and
$2.0$~meV for $a^e_z=1.5$~nm. Also the high-energy part of the spectral
density changes and becomes extended to larger energies. This widening
reflects the fact that the wave functions are now narrower in the
$z$-direction. Note that on the low-energy side the spectral density is the
same for all three curves, because here the behavior is dominated by the
cubic dependence resulting from the coupling matrix element and the phonon
density of states, reflecting the super-ohmic nature of the carrier-phonon
coupling.

We are now looking for a spherical QD, which gives the same spectral density
as for an arbitrary lens-shaped dot. Because the phonon spectral density is a
one-dimensional quantity depending only on the frequency $\omega$, it is
possible to reproduce an arbitrary phonon spectral density by a spherically
shaped QD even with equal localization lengths of electrons and holes. To show this, we assume equal and isotropic form factors $F^{e}_{q}=F^{h}_{q}=F_{q}$. Then the spectral
density reads
\begin{eqnarray}
 J(\omega) &=& \sum_{\bf q} \frac{q}{2V\rho\hbar c_{LA}} (D_e-D_h)^2 F_q^2
 \delta(\omega-\omega_{\bf q})  \notag \\
		  & =& \frac{\omega^3}{4\pi^2 \rho \hbar c_{LA}^5} (D_e-D_h)^2 F_{\frac{\omega}{c_{LA}}}^2.
\end{eqnarray}
Taking the spectral density $J(\omega)$ of an arbitrary QD as given, we thus obtain a spherically symmetric form factor
\begin{eqnarray}
 F_{q} &=& \left[ \frac{4\pi^2 \rho
 \hbar c_{LA}^2}{q^3 (D_e-D_h)^2}J(c_{LA}q)\right]^{1/2} \notag
\end{eqnarray}
which reproduces exactly the given spectral density. Using Eq.~\eqref{eq:ff_wavefunction}, we obtain the square modulus of the
corresponding wave function by inverse Fourier transformation via
\begin{equation}
	|\psi(r)|^2=\frac{1}{(2\pi)^3}\int d^3q \,F_{q} \, e^{-i\mathbf{q}\cdot\mathbf{r}}
= \int\limits_{0}^{\infty} \frac{q \sin(qr)}{2\pi^2 r} \, F_q \, dq \notag
\end{equation}
Note that we here took into account already the spherical symmetry.
Considering the ground  state, the wave function does not have nodes and we
can assume $\psi(r)=\sqrt{|\psi(r)|^2}=r^{-1} u(r)$. The function $u(r)$
satisfies a radial Schr\"odinger equation
\begin{eqnarray}\label{eq:exp_SG}
	\left( -\frac{\hbar^2}{2m_e} \frac{d^2}{dr^2} + V(r)\right) u(r) = E\,u(r)
\end{eqnarray}
with an effective potential $V(r)$. Setting the energy $E$ to zero, the
potential follows from
\begin{eqnarray} \label{eq:exp_pot}
	V(r) =\frac{\hbar^2}{2m_e}\frac{1}{u} \frac{d^2 u}{dr^2}.
\end{eqnarray}

Equation~\eqref{eq:exp_pot} allows us to calculate for a given phonon
spectral density of a non-spherical QD a fictitious spherically symmetric
effective potential. By construction, the wave function of this fictitious QD
exactly reproduces the phonon spectral density of the original non-spherical
QD. It should be noted that the choice of the electron mass $m_e$ in
Eqs.~(\ref{eq:exp_SG}) and (\ref{eq:exp_pot}) is arbitrary; one might also
choose the hole mass or any other mass, which would simply scale the
potential.

\begin{figure}[t]
	\includegraphics[width=\columnwidth]{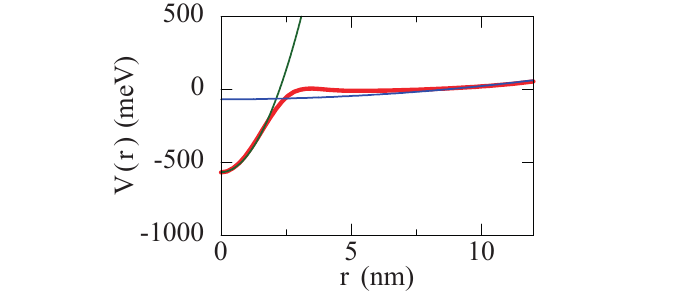}
	\caption{\label{fig:pot} Spherically symmetric effective potential (red line) obtained from
Eq.~(\ref{eq:exp_pot}) for a lens-shaped QD with $a^e_r=5$~nm and $a^e_z=1.5$~nm.
For comparison the parabolic electron potentials are shown for the
in-plane (blue line) and out-of-plane (green line) direction.}
\end{figure}

Already for lens-shaped dots with harmonic confinement, an analytical
solution of Eq.~(\ref{eq:exp_pot}) is lengthy and not very instructive.
Figure~\ref{fig:pot} shows the numerically calculated effective potential
$V(r)$ reproducing the phonon spectral density of the lens-shaped QD with
$a^e_r=5$~nm, $a^e_z=1.5$~nm, and $a^h/a^e=0.87$ of
Fig.~\ref{fig:spectraldensity}(a) (blue curve). The thick red line is the
effective potential calculated by the procedure outlined above. Obviously,
the exciton potential is not anymore harmonic, but is essentially composed of
two parts that exhibit a harmonic shape. For comparison, the in-plane (blue)
and out-of-plane (green) potentials of the lens-shaped QD are marked by thin
lines. (Note that the potentials for electrons and holes are the same.) For
$r\to 0$ the effective potential is dominated by the potential in
$z$-direction showing a steep harmonic behavior. For larger distances the
in-plane harmonic potential dominates the effective potential. In between the
potential is continuous with a local maximum around the region where the
harmonic potentials cross.

While the procedure described above allows us to construct a spherically
symmetric QD which exactly reproduces the phonon spectral density of an
arbitrarily shaped QD, we will now show that for lens-shaped QDs with
harmonic confinement potentials the spectral density can even be well
approximated by spherically symmetric Gaussian wave functions for electron
and hole. In the following we will use the most strongly lens-shaped QD in
Fig.~\ref{fig:spectraldensity}(a) (blue curve) as a reference and call this
QD A. In Fig.~\ref{fig:spectraldensity}(b) we compare this QD with two spherically symmetric ones, QD B and QD C.
The parameters of these three dots are
\begin{itemize}
	 \item[] \textbf{QD A}: $a^e_r=5.0$~nm, $a^e_z=1.5$~nm, $a^h/a^e=0.87$
 	 \item[] \textbf{QD B}: $a^e_r=a^e_z=3.6$~nm, $a^h/a^e=0.87$ 	
 	 \item[] \textbf{QD C}: $a^e_r=a^e_z=4.7$~nm, $a^h/a^e=0.40$	
\end{itemize}
QD B is a  spherical dot with the standard localization length ratio $a^h /
a^e = 0.87$ and a radial localization length $a^e=3.6$~nm chosen in such a
way that it has the same Huang-Rhys factor
\cite{krummheuer2002the,huang1950the,duke1965pho} as QD A, i.e., the same
value for
\begin{equation}
S = \int \frac{J(\omega)}{\omega^2} d\omega .
\end{equation}
The Huang-Rhys factor determines the polaron shift and is a measure for the
total coupling strength. 

Fig.~\ref{fig:spectraldensity}(b) shows that the agreement is not very satisfying.
While for energies up to the maximum the spectral density is well reproduced,
this spherical QD completely fails in reproducing the high-energy tail of the
spectral density.

However, when we allow the confinement ratio $a^h/a^e$ to vary, we find a
rather good agreement, as shown for QD C using $a^e=4.7$~nm and
$a^h/a^e=0.40$. This QD, modeled by just two isotropic Gaussian wave
functions, is able to reproduce both the rise of the spectral density to the
maximum at $2.0$~meV and the long tail up to around $8$~meV. Only small
differences in the high-energy tail of the spectral density between QD A and QD C
are visible.
\section{Phonon influence on the electronic system}
\begin{figure}[t]
	\includegraphics[width=\columnwidth]{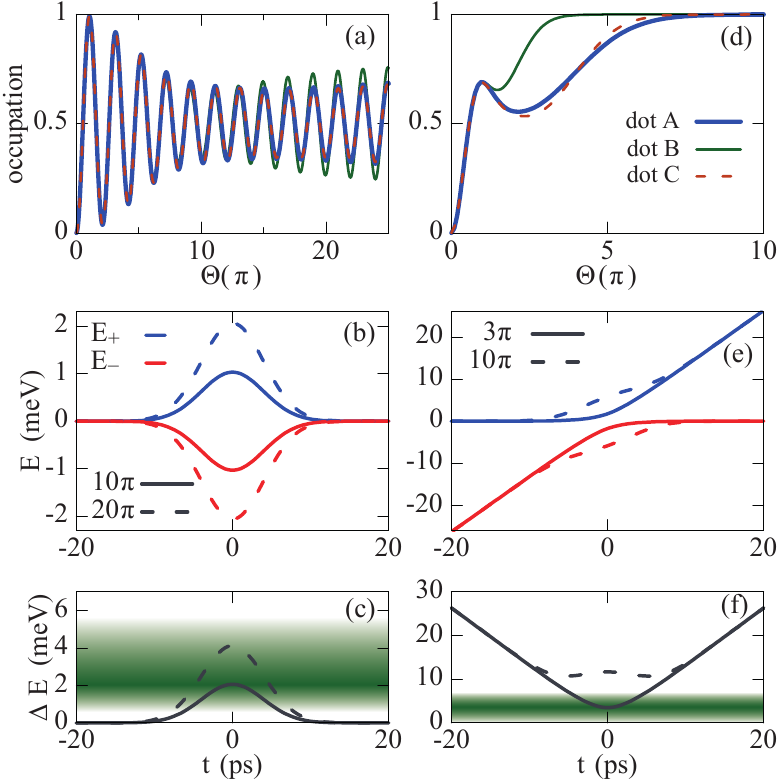}
	\caption{\label{fig:ARP} (a) Final occupation of the exciton
state as a function of the pulse area for resonant excitation with a $4$~ps Gaussian laser
pulse for dot A (blue line), dot B (green line), and dot C (red dashed line). (b)
Instantaneous eigenenergies $E_+$ and $E_-$ for excitation with a resonant Gaussian
$10\pi$-pulse (solid lines) and $20\pi$-pulse (dashed lines). (c) Energy splitting
$\Delta E = E_+ - E_-$ of the eigenenergies for the $10\pi$-pulse (solid line)
and $20\pi$-pulse (dashed line); the green-shaded area highlights the strength
of the phonon spectral density as a function of the energy $\Delta E$.
(d), (e) , and (f): Same as left panel,
but for excitation with a negatively chirped laser pulse with $\alpha = -0.5\,$ps$^2$
and $\tau_0 = 80\,$fs. In (d) and (f) the curves refer to pulse areas of $3\pi$ (solid lines) and $10\pi$ (dashed lines).}
\end{figure}
In the next step, we compare the influence of the QD shape on the optically
excited  exciton occupation of a QD using the lens-shaped QD A and the
spherical QDs B and C. Note that a spherical QD confined by the effective
potential calculated via Eq.~\eqref{eq:exp_pot} exactly reproduces the
results of the lens-shaped QD A (not shown).

The standard example for optical excitations are Rabi rotations, which
recently have been measured for a pulse area up to $12\pi$
\cite{ramsay2010dam,ramsay2010pho}. Here we consider the excitation with a
Gaussian laser pulse described by the instantaneous Rabi frequency (pulse
envelope)
\begin{equation}\label{eq:res_pulse}
\Omega(t)=\frac{\Theta}{\sqrt{2\pi}\tau} \exp\left(-\frac{t^2}{2\tau^2}\right)
\end{equation}
with pulse area $\Theta$, pulse length $\tau=4$~ps, and frequency $\omega$
being resonant with the exciton transition. In Fig.~\ref{fig:ARP}(a) the Rabi
rotations for the three QDs A, B and C are shown. For pulse areas up to about
$12\pi$ we observe a damping of the Rabi rotations with increasing pulse
area, which is essentially the same for all three QDs. Subsequently, the
amplitude of the rotations increases again, which is more pronounced for QD
B, while QDs A and C remain to be in almost perfect agreement. This growth of
the amplitude indicates the regime of the reappearance of Rabi rotations
which has been theoretically predicted \cite{vagov2007non,glassl2011inf}
however not yet clearly seen experimentally, mainly because of the very high
pulse areas necessary to enter this regime.

More insight in the pulse area dependence of Rabi rotations for the different
QDs can be obtained in the dressed state picture. The dressed states are the
eigenstates of the coupled QD-light system
\cite{reiter2014the,wigger2014ene}. In the dressed state picture phonons give
rise to transitions between the dressed states
\cite{mccutcheon11coh,wigger2014ene,kaldewey2017dem} if the actual splitting
is comparable to the energy range covered by the phonon spectral density. The
dressed states as function of time for the Gaussian excitation with a pulse
area of $10\pi$ and $20\pi$ are presented in Fig.~\ref{fig:ARP}(b), while
Fig.~\ref{fig:ARP}(c) shows the energy splitting $\Delta E = E_+ - E_-$. The
green-shaded area highlights the strength of the phonon spectral density of
QD A as a function of the energy $\Delta E$ with its maximum at about
$2$~meV. For the Gaussian pulse the splitting also takes a Gaussian curve
with a maximum at the time of the pulse maximum. Before and after the laser
pulse the eigenenergies are degenerate. Hence, phonons from a large energy
spectrum starting from $0$ up to the maximal splitting contribute to the
damping. For a pulse area of $10\pi$ the splitting reaches the maximum value
of the spectral density [see Fig.~\ref{fig:ARP}(c)]. Therefore, up to such
pulse areas the damping of the Rabi rotations increases. Up to the maximum
the phonon spectral densities of the three dots are very similar leading also
to a very similar behavior of the Rabi rotations. For a pulse area of $20\pi$
the splitting reaches values where the spectral density is already
considerably reduced compared to its maximal value leading to the observed
increased amplitude of the Rabi rotations. In this energy range the spectral
density of QD B is much smaller than for QDs A and C explaining why QD B
exhibits larger Rabi rotations than the other dots. However, from
Fig.~\ref{fig:ARP}(c) we also note that even if the energy splitting at the
pulse maximum is already far beyond the maximum of the spectral density, due
to the degeneracy before and after the laser pulse there is always a time
where the energy splitting of the dressed states passes by the maximum of the
phonon spectral density, which yields the main contribution to the
phonon-induced damping of the Rabi rotations.

The second example is an excitation using a chirped laser pulse
\cite{wu2011pop,simon2011rob}, which turned out to be very sensitive to the
phonon influence and for which excellent agreement between theory and
experiment has recently been shown \cite{kaldewey2017coh,kaldewey2017dem}. It
is well established that at low temperatures only negatively chirped pulses
exhibit a damping due to phonons
\cite{luker2012inf,mathew2014sub,kaldewey2017coh,kaldewey2017dem}, hence we
concentrate on this case here. The theory for the description of the chirped
pulses can be found, e.g., in \cite{luker2012inf,kaldewey2017dem}. Following
Ref.~\cite{kaldewey2017dem}, we take a resonant Gaussian pulse
[Eq.~(\ref{eq:res_pulse})] with initial pulse length of $\tau_0=80$~fs and
apply a chirp filter with coefficient $\alpha=-0.5$~ps$^2$. As a result the
central frequency of the laser pulse changes linearly with time, while the
pulse length is stretched to $\tau = 6.1\,$ps, the pulse envelope being given
by
\begin{equation}\label{eq:chirp_pulse}
\Omega(t)=\frac{\Theta}{\sqrt{2\pi\tau\tau_0}} \exp\left(-\frac{t^2}{2\tau^2}\right).
\end{equation}
The resulting exciton occupations are shown in Fig.~\ref{fig:ARP}(d) for QDs
A, B and C.

For QD A we find that the occupation rises to a maximum at a pulse area
around $1\pi$.  Then, the damping sets in resulting in a minimum of the
occupation around $2.5\pi$. After the minimum the occupation rises again and
reaches values up to one, i.e., almost perfect exciton generation, showing
clearly the reappearance of the ARP effect \cite{kaldewey2017dem}. The
spherical QD C reproduces this behavior well, but some slight deviations from
the lens-shaped QD A are seen, which can be attributed to the small
differences in the spectral density: Around the minimum the occupation of dot
C is slightly below the occupation of dot A, which is a result of the
mismatch of the phonon spectral densities around $4\,$meV that promotes a
slightly stronger phonon coupling of dot C. On the other hand, the phonon
spectral density of dot A exceeds the one of dot B for energies above
$6\,$meV, which leads to a slightly stronger coupling of high energy phonons
that slightly reduces the recovery of the occupation to $1$ of dot A for
$\Theta > 5\pi$.

Dot B is not able to reproduce the found behavior at all, instead the minimum
is at a  much lower pulse area and also the reappearance occurs at much lower
pulse areas. The behavior can again be understood by looking at the dressed
states. Due to the transformation to the frame rotating with the
instantaneous frequency of the pulse, the energy of one of the eigenstate
changes linearly with time while the other remains fixed. The light pulse now
leads to an avoided crossing of the two states resulting in a finite
splitting between the dressed states at any time [Fig.~\ref{fig:ARP}(d)].
Indeed, due to this fact the low-energy phonons below the minimal splitting
(determined by the amplitude of the light at the pulse maximum) cannot
contribute to the damping of the ARP at all and the phonon impact is
restricted to the high-energy tail of the phonon spectral density. Comparing
this to Fig.~\ref{fig:spectraldensity}(b), where we indeed found the largest
difference for higher energy phonons, it is clear, why the ARP is much more
sensitive to the QD shape than the Rabi rotations. In Fig.~\ref{fig:ARP}(f)
the energy splitting for pulse areas of $3\pi$ and $10\pi$ is plotted
together with the energy dependence of the phonon spectral density (green
shaded area). We clearly see that for the $10\pi$-pulse the energy region of
a non-vanishing phonon spectral density is never reached, explaining why here
we obtain an essentially perfect exciton generation via the ARP process. The
main reason for the different behavior of resonant and chirped excitation in
the reappearance regime is thus the fact that for resonant excitation the
splitting starts from zero and always passes through the region of strong
phonon coupling while for chirped excitation the splitting starts from very
large values and, for sufficiently strong pulses, never reaches the region
where phonon coupling is efficient.

Altogether, the dot shape has an impact on the carrier-phonon coupling.
However, the  decisive quantity that describes the phonon coupling is the
phonon spectral density. Mathematically speaking, the phonon spectral density
is a one-dimensional function depending only on the frequency, where
information about the full three-dimensional phonon coupling is lost. Hence,
it is always possible to find a spherical dot that has the same or a similar
phonon spectral density compared to a more realistic lens-shaped QD. Often
this is even possible by using rather simple wave functions obtained, e.g.,
by using the localization length ratio $a^h / a^e$ as a fit parameter. This
can be very beneficial by considerably reducing computational costs. This is
also of importance when considering systems with more than two states
involving, e.g., biexcitons. \\
We remark that our considerations also hold for other electron-phonon coupling mechanisms like the Fr\"ohlich coupling to optical phonons and the piezoelectric coupling to acoustic phonons. Also in these cases the spectral density can be reproduced by a spherically symmetric model. While in the case of Fr\"ohlich coupling the bulk coupling matrix element is again isotropic, such that the formalism can be directly transferred, in the piezoelectric case already the bulk coupling matrix element is anisotropic. Here, introducing an isotropic model requires in addition a suitable angular averaging of the bulk coupling matrix element.

\section{Phonon dynamics}
We now turn to the dynamics of the phononic system. Though a lens-shaped QD
can be  substituted by an adapted spherical dot to calculate the carrier
dynamics, the phonons generated during the optical control are largely
affected by the geometry of the QD. The phonon properties nowadays receive
much attentions \cite{vanacore2017ul}, also because new schemes make use of
phonons to control QDs. For example, the optical output of a QD can be
controlled by surface acoustic waves \cite{schulein2015fou,villa2017sur} or
the lasing properties of QDs can be modulated by strain waves
\cite{czerniuk2017pic}. As an example of the influence of the QD shape on the
phonon dynamics we discuss the properties of the phonon wave packets emitted
during the optical excitation of a QD
\cite{wigger2014ene,wigger2013flu,krummheuer2005pur}. For that purpose we
consider the relative volume change, i.e., the divergence of the mean
displacement field $\langle \mathbf{u}(\mathbf{r}) \rangle$,
\begin{eqnarray}
\frac{\delta V}{V}(\mathbf{r}) &=& \mathrm{div} \langle \mathbf{u}(\mathbf{r}) \rangle \nonumber \\
&=& \sum_{\mathbf{q}}\sqrt{\frac{\hbar}{2\rho V
\omega_{\mathbf{q}}}}\left(\langle b_{\mathbf{q}}\rangle e^{i\mathbf{q}\cdot\mathbf{r}}
+ \langle b^\dag_{\mathbf{q}}\rangle e^{-i\mathbf{q}\cdot\mathbf{r}}\right),
\end{eqnarray}
which describes the change of the lattice unit cells due to the coherent phonon modes.
\begin{figure}[t]
	\includegraphics[width=\columnwidth]{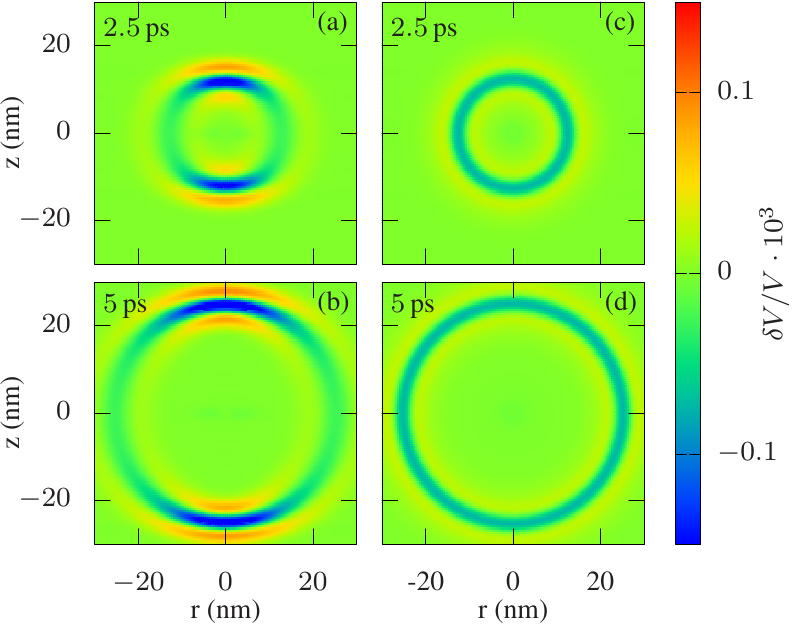}
	\caption{\label{fig:phonon} Relative volume change $\delta V/V$
after an excitation with a $2\pi$-pulse for (a), (b) for the lens-shaped QD A
and (c), (d) for the spherical dot QD C. The upper panels (a),
(c) show the wave packet at $t = 2.5\,$ps and the lower panels (b), (d) at an advanced
time $t = 5\,$ps. }
\end{figure}
Figure \ref{fig:phonon} shows the relative volume change of the lattice after
the optical  excitation of the QD by a Gaussian pulse for the lens-shaped QD
A (left) and the spherical QD C (right). The pulse length is chosen to
$0.5$~ps, which enables an efficient coupling to the phonons
\cite{wigger2014ene}. For the pulse area we take a $2\pi$-pulse, resulting in
the creation and immediate annihilation of the exciton. Whenever an exciton
is present in the dot, the surrounding lattice atoms react on the changed
charge configuration by the formation of a polaron, i.e., a static lattice
deformation. Because we consider long times after the excitation and
de-excitation of the exciton, the polaron has already vanished. However, the
rapid creation and destruction of the polaron leads to the emission of a
phonon wave packet. The details of the phonon emission can be understood by
considering the formation of the polaron \cite{wigger2014ene}.

When the polaron is created a contraction of the lattice unit cells in the QD
region occurs,  i.e., a negative relative volume variation. The strain that
comes along with this contraction is transferred to surrounding lattice
cells, which are in turn stretched. Hence, the leading edge of the emitted
phonon wave packet features a positive volume variation. Afterwards, a front
of a negative volume variation follows that indicates a lattice contraction
similar the polaron. Finally, the trailing edge of the phonon wave packet
exhibits again a positive volume variation that is a fingerprint of the
destruction of the polaron, i.e., the lattice atoms in the QD area return
back to their initial positions, so that the lattice unit cells are stretched
again.

Both the spherical and the lens-shaped QD exhibit the same sequence of
positive and negative  volume variation. However, the emitted wave packet
reproduces the symmetry of the QD. In the case of a lens-shaped QD the symmetry is broken, so that the preferred direction points towards the smallest extension
of the dot. Accordingly, in Fig.~\ref{fig:phonon}(a), (b) we observe the
emission of a  phonon wave packet, but it is clearly visible that the
amplitude is much stronger in $z$-direction than in the in-plane direction.
This can be related to the coupling strength for the phonons, which becomes
stronger the smaller the confinement is. For the spherical QD in Fig.~\ref{fig:phonon}(c), (d) we see a ring moving outwards, reflecting the spherical shape of the QD. We note that for the case of piezoelectric coupling the anisotropic bulk coupling matrix element introduces an additional anisotropy in the phonon emission besides the anisotropy introduced by the QD geometry \cite{krummheuer2005pur}. For the Fr\"ohlich coupling to optical phonons, on the other hand, an anisotropic polaron will build up for an antisymmetric QD, however, due to the vanishing group velocity no phonon wave packet will be emitted into the surrounding material.

The usage of stronger pulses or
sequences of pulses can result in the creation of wave packet trains
\cite{wigger2014ene} or even squeezed phonons \cite{wigger2013flu}, which
would be emitted into a direction that is determined by the QD geometry. This
shows that by tailoring the QD geometry one gains control over the phonon
generation.

\section{Conclusion}
By explicitly comparing the electron-phonon coupling of a spherical QD and a
lens-shaped QD  using the same theoretical approach we draw two conclusions:
For the electronic system, it is sufficient to use the spherical symmetry
even if the QD shape is more involved. This can be traced back to the fact
that the decisive quantity in this case is the phonon spectral density, which is
a one-dimensional function, and hence can be exactly reproduced by a radial
QD symmetry that is adapted to the properties of the lens-shaped dot. For the
phonon system, the actual shape plays an important role in determining the
properties of the created phonons. A directed emission of phonons can only be
modeled by including the actual shape of the QD. The strongest emission is
along the smallest axis of the QD. By performing a direct comparison, we shed
new light on the question regarding the influence of the QD geometry on the
optical state preparation of self-assembled QDs.

\begin{acknowledgments}
We thank Daniel Wigger and Martin Axt for fruitful discussions.
\end{acknowledgments}

\end{document}